\documentstyle[prl,aps,epsf,twocolumn]{revtex}
\begin{document}
\title{ 
One-Dimensional Stochastic L\'evy--Lorentz Gas.}
\author{E. Barkai$^a$, V. Fleurov$^a$ and J. Klafter$^b$  \\
$^a$ School of Physics and Astronomy \\
$^b$ School of Chemistry \\
Beverly and Raymond Sackler Faculty of Exact Sciences\\
Tel-Aviv University\\
Tel-Aviv 69978, Israel}
\date{\today}
\maketitle

\begin{abstract}

 We introduce a L\'evy--Lorentz gas 
in which a light particle is scattered by static point
scatterers arranged on a line. We investigate the case where
the intervals between scatterers $\{ \xi_i \}$ are independent
random variables
identically
distributed 
according to the probability
density function $\mu\left( \xi \right)\sim \xi^{-\left( 1 + \gamma\right)}$.
We show that under  certain conditions 
the mean square displacement of the particle obeys
$\langle x^2 \left( t\right) \rangle \ge C t^{3 - \gamma}$ for $1 < \gamma < 2$.
This behavior is compatible with 
a renewal L\'evy walk scheme. 
We discuss the importance of rare events in the proper characterization
of the diffusion process.

$$ $$

\end{abstract} 





\section{Introduction}

 In recent years 
there has been
a growing interest in anomalous diffusion
defined by 
\begin{equation}
\langle x^2  \rangle = D_{\delta} t ^ \delta
\label{eq000}
\end{equation}
and $\delta > 1$ 
\cite{Bouch,Levy,Klafter1,Benkadda}.  
Such a behavior was found
in chaotic diffusion in low dimensional 
systems \cite{Geisel,Klafter5}, tracer diffusion in a rotating
flow \cite{Swin}, $N$ body Hamiltonian dynamics \cite{Antony},  Lorentz
gas with infinite horizon \cite{Bouch,Mats}  and
diffusion in egg crate potentials \cite{Geisel2}.
In all these examples one observes long ballistic
flights in which the diffusing particle
moves at a constant velocity. 
The transport is characterized by a distribution
of free flight  times  which follows a power law
decay. These processes have been usually analyzed using 
the L\'evy walk framework (see more details below)
\cite{Levy,Klafter1,Benkadda,Klafter5}.

 It has been recently suggested by Levitz \cite{Levitz} that three-dimensional
molecular Knudsen diffusion, at very low pressures,
inside porous media
can be described by L\'evy walks.
It has been also shown 
\cite{Levitz1} that pore chord distributions,
in certain
three-dimensional porous media 
decay as a power law,
at least for several length scales.
Hence one can anticipate that a light test particle 
injected into such a medium may exhibit 
a L\'evy walk.
This has motivated the investigation of a fractal
Lorentz gas.
Levitz \cite{Levitz}  has simulated  trajectories
of  a light particle
reflected from a three-dimensional intersection of a 
four-dimensional Weierstrass--Mandelbrot hyper surface,
and found an enhanced L\'evy type diffusion.

Here we investigate a one dimensional stochastic
Lorentz gas
which we call L\'evy--Lorentz gas.
In this model a light particle is scattered by a fixed
array of identical scatterers arranged randomly on a line.
Upon each collision event the light particle can be
transmitted (or reflected) with probability $T$ (or $R=1-T$). 
We investigate the case when the intervals between the scatterers
are independent identically distributed random variables
with a diverging variance.

 We find: ({\bf a})  a lower bound for the mean square displacement
which is compatible with the L\'evy walk model, and ({\bf b})
that the generalized diffusion coefficient $D_\delta$ is
very sensitive to the way the system has been prepared at
time $t=0$. 
As expected, we  show that the transport
is not Gaussian. 
In systems that exhibit normal diffusion,
the contribution
from ballistic motion, $x^2 = v^2 t^2$, is
important only for short times; here we show that
the ballistic motion cannot be neglected even at $t \to \infty$.
The ballistic paths contribute to the generalized diffusion
coefficient $D_\delta$ exhibiting a  behavior
different than normal.

\section{Model and Numerical Procedure}

\begin{figure}[htb]
\epsfxsize=21\baselineskip
\centerline{\vbox{
      \epsffile{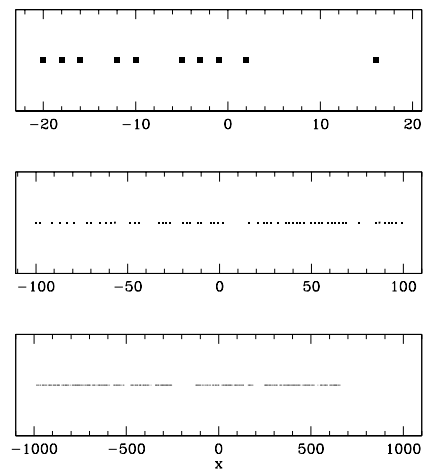}  }}
\caption {
A realization of a set of scatterers with $\gamma=3/2$ exhibiting
gaps on many scales.
The horizontal axis is the $x$
coordinate. All along this work
we consider dimensionless units. 
}
\label{fitLevy0}
\end{figure}

 Assume a light particle which moves with a constant speed
($v=\pm 1$) among identical point scatterers arranged randomly 
on a line. 
Upon each collision, 
the probability that the light particle
is transmitted (reflected)  is $T$ $(R=1-T)$.
The intervals between scattering points,
$\xi_i> 0$ with  $\left( i=\cdots,-n,\cdots,-1,0,1,\cdots\right)$,
are independent identically distributed random variables
described by a probability density function $\mu\left( \xi \right)$.
An important random variable
is $x_f$
defined to be the distance between
the initial location of the light particle $(x=0)$
and the first scatterer in the sequence 
located at $x>0$.
The random variable $x_f$ is described by 
the probability density function $h(x_f)$.
A set of scatterers (black dots) is given schematically by:
 $$ $$
$$ \cdots  \ \ \overbrace{  \bullet  \ \ \ \ \ \    \bullet }^{\xi_{
-2}} \overbrace{ \ \ \ \ \ \  \bullet }^{\xi_{-1}} \overbrace
{  \ \ \ \ \ \underbrace{\circ \ \ \ \  \bullet}_{x_f} }^{\xi_0} \overbrace{  \ \  \  \ \ \  \  \bullet }^{\xi_1}
\overbrace{ \  \ \  \ \ \ \ \ \ \   \ \bullet}^{\xi_2} \ \  \ \cdots $$
$$ $$
where the open circle represents the light test particle at time $t=0$.
We consider the case when
for large $\xi$,
$\mu\left(\xi\right) \sim \xi^{-\left( 1 + \gamma\right)}$,
with $0< \gamma < 2$. Thus the variance of the length intervals
$\{ \xi_i \}$ diverges.
A realization of the scatterers is shown in Fig. \ref{fitLevy0},
for the case $\gamma=3/2$.
We observe large gaps which are of the order
of the length of the system.

 The case for which the variance converges
has been investigated thoroughly
in \cite{Henk,Gras,Spohn,Ernst,barkaiJSP}, resulting in:
$(i)$
a normal Gaussian diffusion 
as expected from the central limit theorem,
and  $(ii)$
a $3/2$  power law decay in $t$ of
the  velocity autocorrelation function. 
 
 Along this work we present numerical results for the
case $\gamma=3/2$ and $T=1/2$.
We use the following numerical procedure.
First we  generate a set of scatterers 
on a one dimensional lattice with a lattice
spacing equal unity.
Using a discrete time
and space iteration scheme we find an exact expression
for 
the probability of finding the particle on $x$
at time $t$, 
$p(x,t|x=0,t=0)$, given that at $t=0$
the particle is on  $x=0$.
The initial location of the particle is determined
using equilibrium initial conditions (see details below).
The initial velocity is $v=1$ or $v=-1$ with equal 
probabilities.
$p(x,t|x=0,t=0)$ depends on the realization of disorder
we have generated in the first step. 
This procedure is repeated many times.

In appendix A we explain how we generated
the random intervals $\{\xi_i\}$. 
When $\gamma=3/2$ the
mean $\langle \xi \rangle \equiv \int_0^{\infty} \mu(\xi) d \xi$
is finite while the second moment
$\langle \xi^2 \rangle=\infty$. 
Since
$|v| = 1$
the characteristic microscopic time scale is $\langle \xi \rangle$
which is referred to as the mean collision time,
and our simulations are for times $t \sim 1000 \langle \xi \rangle$.
For our
choice of parameters $\langle \xi \rangle \sim 4$ (see more details
in Appendix A). 

\section{Results}

 Let us analyze our one-dimensional L\'evy--Lorentz model using
the  L\'evy walk approach
\cite{Levy,Klafter1,Benkadda,Klafter5,previous4}.
L\'evy walks describe random walks which 
exhibit enhanced diffusion and are based on the  generalized
central limit theorem and L\'evy stable distributions \cite{Feller}.
Briefly, a particle moves with a constant velocity
$v=+1$ or $v=-1$  and then at a random time $\tau_1$ its velocity is changed.
Then the process is renewed. Each collision is independent of
the previous collisions. The 
times between  collision events 
$\{ \tau_i \}$
are assumed to be independent identically distributed random
variables,
given in terms of a probability density function $q\left( \tau \right)$.
One might expect that the dynamics of L\'evy--Lorentz gas
can be analyzed using the L\'evy walk  renewal approach 
with
$q\left( \tau \right) \sim \tau^{ -\left( 1 + \gamma\right)}$,
for large $\tau$ and $0< \gamma < 2$,
which leads to 
\begin{equation}
\langle x^2 \rangle \sim \left\{ 
\begin{array}{cc}
t^{3 - \gamma}, &  \ \ \   1< \gamma < 2 \\
 \  &  \ \ \   \\
t^{2}, & \ \ \  0< \gamma < 1. 
\end{array}
\right.
\label{eqLL02}
\end{equation}
For $\gamma>2$ one finds normal diffusion.
It is clear that the renewal L\'evy walk approach and the
L\'evy--Lorentz gas are very different. Within the L\'evy--Lorentz gas
collisions are not independent and correlations are important.
Hence it is interesting to check whether the renewal L\'evy
walk model is suitable for the description of the L\'evy--Lorentz
gas.
In this context it is interesting to recall that Sokolov 
et al  \cite{Sokolov}
have shown that correlations between jumps
in a L\'evy flight in a chemical space 
destroy the L\'evy statistics of the walk.

%
%
%

\begin{figure}[htb]
\epsfxsize=21\baselineskip
\centerline{\vbox{
      \epsffile{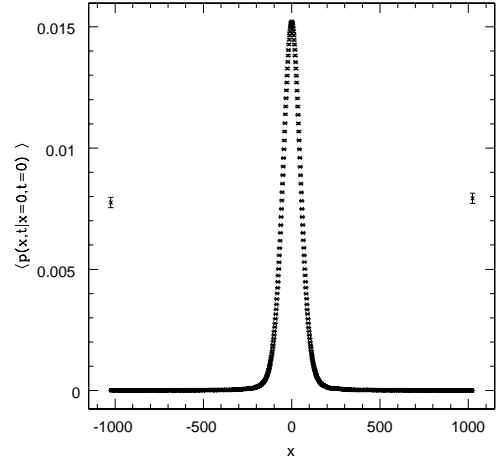}  }}
\caption {
A histogram presenting the 
$\langle p(x,t|x=0,t=0) \rangle$ versus $x$
for time $t=1024$,  $\gamma=3/2$
and $T=1/2$.
Notice the ballistic peaks of the propagator at $x=\pm t$.
The average is over $1.8*10^5$ realizations of disorder.
The bin length is unity.
}
\label{FIG.pxens1}
\end{figure}

\begin{figure}[htb]
\epsfxsize=20\baselineskip
\centerline{\vbox{
      \epsffile{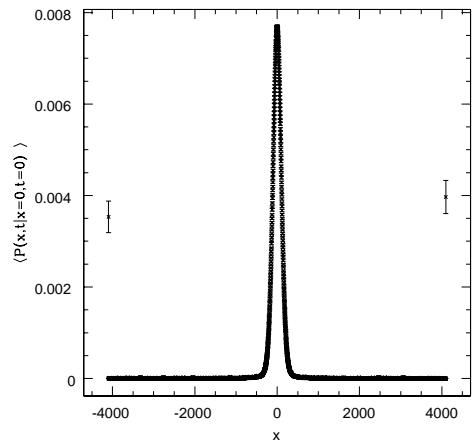}  }}
\caption {
The same as Fig. \protect\ref{FIG.pxens1}
for time $t=4096$. The probability 
of finding a ballistic path, 
$\langle
p(x=\pm t, t |x=0,t=0)\rangle \simeq 0.004$,
is small but yet of statistical
significance when $\langle x^2 (t) \rangle$ is calculated.
}
\label{fitLevy1}
\end{figure}

We consider a continuum model to derive
our analytical results; the generalization to
the lattice case is straightforward.
 Let $\langle p\left( x , t | x=0, t = 0 \right) \rangle dx$
be the probability, averaged over disorder, of finding
the test particle at time $t$, in the interval $(x,x+dx)$.
Initially, at  time $t=0$, the particle is
at $x=0$, and there is an equal probability
of the particle having a velocity $v=+1$ or $v=-1$.
Figs. \ref{FIG.pxens1} and \ref{fitLevy1} present numerical simulations
which show
$\langle p\left( x , t | x=0, t = 0 \right) \rangle$.
One can see that in addition
to the central peak on $x=0$, two other peaks
appear at locations
$x=\pm t$. These peaks, known as ballistic
peaks, were observed in a similar context
in other systems exhibiting enhanced diffusion 
\cite{Klafter5,Levitz,West}. The peaks are stable on the
time scale of the numerical simulation.
 The height of these peaks decays with time, 
and according to our finite time numerics
the central peak and  the ballistic peaks
decay according to the same power law when
$\gamma=3/2$.

Let us analyze analytically the time dependence of
the  ballistic peaks and calculate
their contribution to the mean square displacement.
 Since in our model $|v|=1$ it is clear that 
\begin{equation}
 \langle p\left( x , t | x=0, t = 0 \right) \rangle = 0, \ \ \ \mbox{for} \ \ |x|> t. 
\label{eqLL03}
\end{equation}
We decompose the ensemble averaged probability density into 
two terms
 $$ \langle p\left( x , t | x=0, t = 0 \right) \rangle =  $$
\begin{equation}
 \langle \tilde{p}\left( x , t | x=0, t = 0 \right) \rangle + 
{1 \over 2} Q_b\left( t \right) \left[  \delta\left(x+t \right) + \delta\left(x-t\right) \right].
\label{eqLL04}
\end{equation}
The first term on the RHS,
 $\langle \tilde{p}\left( x , t | x=0, t = 0 \right) \rangle$,
  is the probability density of finding
the light particle at $|x|< t$. $Q_b\left( t \right)$ is the 
probability of finding the light particle at $x=t$
($x=-t$) if initially  
at $x=0$ and its velocity
is
$+1$ ($-1$). 
The left-right symmetry in Eq. (\ref{eqLL04})
means that we have used the symmetric
initial condition (i.e., $v=+1$ or $v=-1$
with equal probabilities) and the assumption that the
system of scatterers is isotropic in an averaged sense.
Using a similar notation we write
\begin{equation}
\langle x^2 \rangle = \langle \tilde{x}^2 \rangle + 
\langle x^2 \rangle_b,
\label{eqLL05}
\end{equation}
where $\langle x^2 \rangle_b$ is the ballistic contribution
to the mean square displacement.
From Eq. (\ref{eqLL04}) we have
\begin{equation}
Q_b\left( t \right) t^2 \le \langle x^2\left( t \right) \rangle \le t^2.
\label{eqLL06}
\end{equation}
The upper bound is an obvious consequence 
of the fact that $|v|=1$. The lower bound,
found using $\langle x^2(t)\rangle_b \le \langle x^2(t) \rangle$,
 is of no
use when all moments of $\mu\left( \xi \right)$ converge,
since then $Q_b \left( t \right)$ usually 
decays exponentially for long times. 
Eq. 
(\ref{eqLL06}) is useful when the 
moments of $\mu(\xi)$ diverge, a case we consider here.

%
%
%

 To find $Q_b \left( t \right)$ consider a test particle
which is initially of  velocity $+1$
and located at $x=0$. The probability it
reaches $x=t$, at time $t$,
is $T^r$ where $r$ is the number of scatterers
in the interval of length $(0,t)$.
Hence,
\begin{equation}
Q_b \left( t \right) = \sum_{r = 0}^{\infty} T^r G_r \left( t \right),
\label{eqLL07}
\end{equation}
and $G_r \left( t \right)$ is the probability
of finding $r$ scatterers in $(0,t)$.
$G_r \left( t \right)$ 
can be calculated in terms of $\mu\left( \xi \right)$ and of 
$h\left(x_f\right)$.
In appendix B we use renewal theory to calculate
the  Laplace $t \to u$ transform 
of  $Q_b\left( t \right)$
\begin{equation}
\hat{Q}_b\left( u \right) =
 { 1 \over u} + { \left ( T - 1 \right) \hat{h}\left( u \right) \over
\left[ 1 - T \hat{\mu}\left( u \right) \right] u}.
\label{eqLL08}
\end{equation}
When $T=1$, $\hat{Q}_b\left( u \right) = 1/u$, as expected from a
transmitting set of scatterers.
In deriving Eq. (\ref{eqLL08}) we have used the model assumptions
that the intervals $\{ \xi_i \}$ are statistically
independent and identically distributed.

 The function $h(x_f)$ depends on the way the system
of scatterers and light particle are initially
prepared. Consider the following preparation process.
A scatterer is assigned at the location $x=-L$
(eventually $L \to \infty$), then random  independent length intervals
are generated using the probability density $\mu(\xi)$. These
length intervals determine the location of scatterers on the line.
When the sum of the length intervals exceeds $2L$ the
process is 
stopped. As mentioned, at time $t=0$ the light particle is
assigned to the point $x=0$.  
When the mean distance between scatterers
$\langle \xi \rangle = \int_0^{\infty} \xi \mu(\xi) d\xi$
converges (i.e.,  $1< \gamma$), and  $L \to \infty$ then 
according to \cite{Feller,COX}
\begin{equation}
h\left( x_f \right) = { 1 - \int_0^{x_f} \mu(\xi) d\xi \over \langle \xi \rangle},
\label{eqLL09}
\end{equation}
which  is standard in the context of 
the Lorentz gas 
when the moments of $\mu(\xi)$
converge 
\cite{Henk,Spohn}. 
This type of initial condition
is called  equilibrium initial condition. 
 When $1< \gamma < 2$, Eq. (\ref{eqLL09})
implies that
$h(x_f) \sim \left( x_f\right)^{ - \gamma}$ and hence 
$\langle x_f \rangle = \int_0^{\infty} x_f h(x_f) dx_f \to \infty$.
At first sight
this divergence
might seem to be paradoxical, since 
the mean distance between scatterers, 
$\langle \xi \rangle$,
converges.
We notice however  that the point $x=0$ 
has a higher probability to be situated
in a large gap.
Hence, statistically the interval $\xi_0$ is much
larger than  the
others and in our case
$\langle x_f \rangle= \infty$.
Eq. (\ref{eqLL09}) implies that on average
one has to wait an infinite time for the 
first collision event.

%
%
\begin{figure}[htb]
\epsfxsize=20\baselineskip
\centerline{\vbox{
      \epsffile{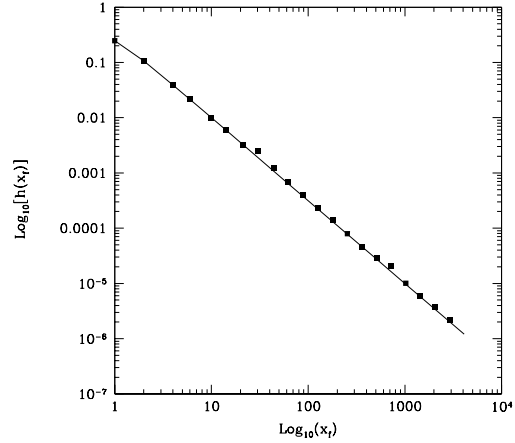}  }}
\caption {
The probability to find the first scatterer
at a distance $x_f$ from the origin. Here the average is over $3*10^5$
realizations, and half the length of the system is $L=10^5$.
We use a bin of length $32$ (dimensionless units).
The solid curve is the theoretical prediction,
Eq. (\protect\ref{eqLL09}),
with no fitting parameters.
For large $x_f$, $h(x_f)\sim x_f^{ -\gamma}$ 
and $\gamma=3/2$,
which implies that  $\langle x_f \rangle$ diverges.}
\label{fitLevy2}
\end{figure}

 In numerical simulations the system's length $L$
is finite, so that  Eq. (\ref{eqLL09}) is only an 
approximation which we expect to be valid for 
$x_f<<L$.
However, if we observe a system for time $t <<L$
the boundary  condition is not expected
to influence the anomalous dynamics.
We have generated numerically many random systems,
using $\mu(\xi) \sim \xi^{ - \left( 1  + \gamma \right)}$ and $\gamma=3/2$.
As shown in  
Fig. \ref{fitLevy2},
$h(x_f) \sim \left( x_f\right)^{ - \gamma}$ as predicted
in Eq. (\ref{eqLL09}).

 We consider the small $u$ expansion, 
of the Laplace transform of $\mu(\xi)$,
\begin{equation}
\hat{\mu}\left( u \right) = 1 -  \langle \xi \rangle u + a \left( \langle \xi \rangle u \right)^{\gamma} \cdots, 
\label{eqLL10}
\end{equation}
where $1 < \gamma < 2$ and $a$ is a constant.
Using a Tauberian theorem and 
Eqs. 
(\ref{eqLL08}), 
(\ref{eqLL09})
it can be shown that for long times $t$
$$ Q_b \left( t \right)=$$
\begin{equation}
 {a \over \Gamma\left( 2 - \gamma \right)} \left({ t \over \langle \xi \rangle}\right)^{1 - \gamma} + {  2 a \left( \gamma - 1 \right)  \over \Gamma\left( 2 - \gamma \right) }{ T \over 1 - T} \left( { t \over \langle \xi \rangle} \right)^{ - \gamma} + \cdots .
\label{eqLL11}
\end{equation}
Inserting Eq. 
(\ref{eqLL11})
in Eq. (\ref{eqLL06}) we find 
\begin{equation}
{a \langle \xi \rangle^2 \over \Gamma\left( 2 - \gamma \right) }\left( { t \over \langle \xi \rangle} \right)^{3 - \gamma} \le \langle x^2 \left( t \right) \rangle \le t^2
\label{eqLL12}
\end{equation}
This bound demonstrates that the diffusion is enhanced, namely the mean square
displacement increases faster than linearly with time.

\begin{figure}[htb]
\epsfxsize=21\baselineskip
\centerline{\vbox{
      \epsffile{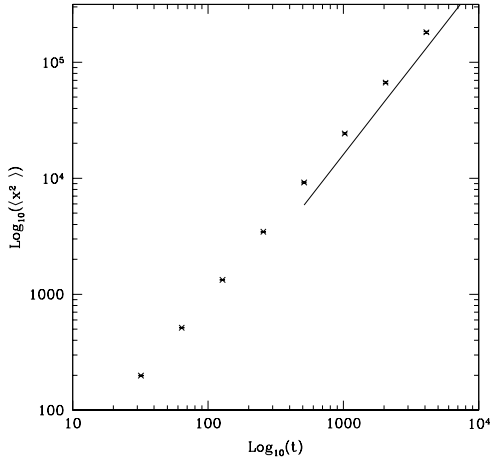}  }}
\caption {
$\log_{10}\left[ \langle x^2 \rangle \right]$ versus $\log_{10}(t)$.
The points are numerical results. The straight curve is the
asymptotic behavior of the lower bound,
Eq. (\protect\ref{eqLL12}) (i.e., $\langle x^2\rangle_b$).
We use $\gamma=3/2$ and so $\langle x^2 \rangle \ge C t^{3/2}$.
}
\label{fitLevy3}
\end{figure}

In Fig. 
\ref{fitLevy3}, we present the mean square displacement
of the light particle obtained by  numerical simulations
for the case $\gamma = 3/2$. We see that for the chosen 
values of parameters the asymptotic $t^{3 - \gamma}$ 
behavior can be observed for times which are accessible 
on our computer. Fig. \ref{fitLevy3} clearly shows
 that the ballistic contribution
$\langle x^2 \rangle_b$ to the mean square displacement
$\langle x^2 \rangle$ is significant.
Notice that our numerical results are
presented for times which are much larger than 
the mean collision time
$\langle \xi \rangle \sim 4$.

\begin{figure}[htb]
\epsfxsize=21\baselineskip
\centerline{\vbox{
      \epsffile{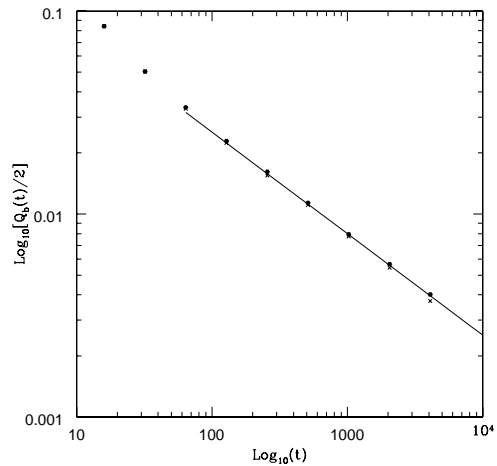}  }}
\caption {
The probability to find the light particle
at time $t$ at $x = + t$ (stars) and at $x=-t$
(dots) versus time.
The solid curve is the theoretical prediction,
Eq. (\protect\ref{eqLL11})
(no fitting parameters) which gives
$Q_b(t)/2 \sim t^{(1-\gamma)}$,
with $\gamma=3/2$. 
}
\label{fitLevy4}
\end{figure}

In Fig. \ref{fitLevy4} we show the probability of
finding a ballistic path, namely,
the probability of finding the light particle at time $t$ at $x=+t$,
or at $x=-t$.
By definition these probabilities are
equal to $Q_b(t)/2$.
We observe the  $t^{1- \gamma}$ behavior of $Q_b(t)$, Eq. 
(\ref{eqLL11}), with which our lower bound was found.
  The fact that 
the probability of finding the particle at $x=t$ is
equal to the probability of finding the particle
at $x=-t$ means that our system
is isotropic in an averaged sense.
This is achieved by choosing large values of $L$. 

%
%
%

 The lower bound in Eq. (\ref{eqLL12}) does not depend on the
transmission coefficient $T$. Thus, even when all
the scatterers are perfect reflectors, with $R=1$, 
the diffusion is enhanced. Large gaps which are of the order of 
the length $t$ are responsible for this behavior. 
 The transmission coefficient has an important role
in determining what is the asymptotic
time of the problem. The condition that the first term
in Eq. (\ref{eqLL11}) dominates over the second reads:
\begin{equation}
{ t \over \langle \xi \rangle }\gg  2 \left( \gamma - 1 \right) { T \over 1 - T}.
\label{eqLL13}
\end{equation}
Only under this condition the behavior in Eq. (\ref{eqLL12})
is expected to be valid. 

The lower bound in  Eq. (\ref{eqLL12}) is compatible with the renewal L\'evy
walk approach Eq. (\ref{eqLL02}).  Other stochastic models
\cite{Bouch,Fog}
for enhanced diffusion based on L\'evy scaling arguments
predict
\begin{equation}
\langle x^2 \rangle \sim t^{ 2 / \gamma} \ \ \mbox{for} \ \ \ 1< \gamma < 2
\label{eqLL14}
\end{equation}
which is different from Eq. (\ref{eqLL02}).
This approach is based upon a fractional
Fokker--Planck equation (FFPE)
\begin{equation}
{\partial p\left( x, t \right) \over \partial t} = D_{\gamma}
 \nabla^{\gamma} p \left(x,  t\right)
\label{eqLL15}
\end{equation}
used in \cite{Fog}
to predict an enhanced diffusion.
The non-local  fractional operator  in Eq. (\ref{eqLL15})
is defined in Fourier $k$ space according
to the transformation $\nabla^{\gamma} \to -|k|^{\gamma}$.
Our findings here show that Eq. (\ref{eqLL14})
does not describe the dynamics of the
L\'evy--Lorentz gas, since $3-\gamma \ge 2/\gamma$
for $ 1\le \gamma \le 2$.

%
%
%

 Consider now the case when  
the light particle is initially located at a scattering point.
Such an initial condition is called non equilibrium initial 
condition.
Under this condition $h(x_f)=\mu(x_f)$ instead of 
Eq. (\ref{eqLL09}).
This means that the particle has to wait  an average
time $\langle \xi \rangle$ before
the first collision event instead of the infinite time
when the equilibrium initial conditions were used.
Using Eqs.
(\ref{eqLL05}), (\ref{eqLL08})
and 
(\ref{eqLL10}),
 and 
$\hat{\mu}( u ) = 1 - \left( A u \right)^{\gamma} + \cdots$
for $0< \gamma < 1$ and small $u$, we find
\begin{equation}
\langle x^2 \rangle \ge \left\{
\begin{array}{cc}
{1 \over 1 - T} a { \left( \gamma - 1 \right) \over \Gamma\left( 2 - \gamma \right)} \langle \xi \rangle^2 \left( { t \over \langle \xi \rangle }\right)^{2 - \gamma}, & \ \ \ \ 1< \gamma < 2 \\
 &  \\
{ 1 \over 1 - T } { \left( 1 - \gamma\right) \over \Gamma\left( 2 - \gamma \right) } A^2 \left( { t \over A } \right)^{2 - \gamma},  & \ \ \ \ 0< \gamma < 1. 
\end{array}
\right.
\label{eqLL16}
\end{equation}
For $1< \gamma < 2$ the bound differs
from the $t^{3 - \gamma}$ found 
in Eq.  (\ref{eqLL12}),
where we chose $h(x_f)$ 
according to  Eq. (\ref{eqLL09}).
Thus the ballistic contribution $\langle x^2 \rangle_b$,
defined in Eq. (\ref{eqLL05}), behaves differently
for the two ensembles even when $t \to \infty$.
This is very different from regular Lorentz gases,
which in the limit $t\to \infty$ are not sensitive to the
choice of $h(x_f)$.

\begin{figure}[htb]
\epsfxsize=21\baselineskip
\centerline{\vbox{
      \epsffile{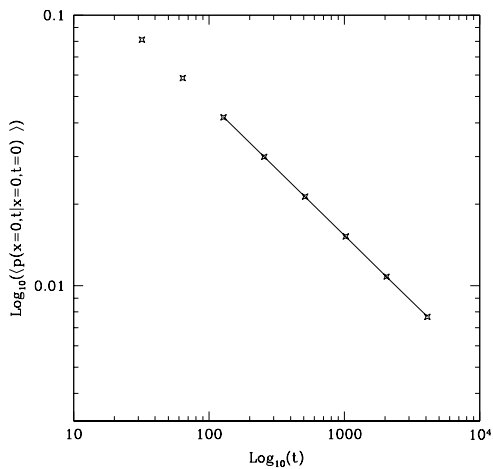}  }}
\caption {
$\log_{10}\left[ \langle p(x=0,t|x=0,t=0) \rangle \right]$ 
versus $\log_{10}(t)$.
The points are numerical results. The straight curve is 
a fit exhibiting the $t^{-1/2}$ behavior.
}
\label{px0}
\end{figure}

 Finally, Fig. \ref{px0} shows the behavior of the correlation
function $\langle p(x=0,t | x=0,t = 0) \rangle$ 
obtained from the numerical
simulation with equilibrium initial conditions.
We observe a
$t^{-1/2}$ decay of the correlation function. This behavior is compatible with
standard Gaussian  diffusion which gives the well known 
$t^{-d/2}$ 
result
in $d$ dimensions. We find this behavior
for time scales which are much larger than the mean
collision time $\langle \xi \rangle$, however we have no proof that this
behavior is asymptotic.  On the other hand the L\'evy walk model
predicts
$\langle p(x=0,t | x=0,t = 0) \rangle\sim t^{-1/\gamma}$
\cite{Klafter5}.

%

%
%
%

\section{Summary and Discussion}

 In this work we have considered a one dimensional
L\'evy--Lorentz gas. 
We have shown that:\\ 
{\bf (a)} the mean square displacement in the L\'evy--Lorentz 
gas 
is compatible with the L\'evy walk framework 
and not with the FFPE. \\
{\bf (b)}  Ballistic contributions to the
mean square displacement are important
even for large times. \\
({\bf c}) The ballistic peaks at $x=+t$ and
$x=-t$ can be analyzed analytically. They decay
as power laws. \\
({\bf d}) The way in which the system
is prepared at $t=0$ (i.e., equilibrium versus
non equilibrium initial conditions) determines the behavior
of the ballistic peaks. Since these peaks contribute
to the mean square displacement even at large times
, we conclude that the diffusion coefficient $D_{\delta}$
is sensitive to the way the system is prepared.


 In our work we considered an initial condition
$v=1$ or $v=-1$ with equal probabilities.
It is clear that if we
assign a velocity $v=+1$ to the light particle,
at $t=0$, $\langle p(x,t|x=0,t=0) \rangle$
will never become symmetric, even approximately. 
Instead of the three peaks in Fig. \ref{fitLevy1} one will
observe only  two peaks one at $x=0$ and the other at $x=+t$.

 The reason for these behaviors in the L\'evy--Lorentz gas
stems from the statistical importance of ballistic paths.
This is different from the systems in which diffusion is normal in which
these paths are of no significance at long times.
Thus, similarly to Newtonian dynamics, the system exhibits
a strong sensitivity to initial conditions.

 Experiments measuring diffusion phenomena 
usually sample  data 
only in a scaling regime (e.g., $-\sqrt{D_1t} < x < \sqrt{D_1t}$
). Rare events where the diffusing particle is
found outside this regime are many times assumed to be of
no statistical importance. Here we showed that for the 
L\'evy--Lorentz gas rare events found in the outer most part
of $\langle p(x,t|x=0,t=0) \rangle$ are of statistical importance.

$$ $$
{\em Note added in proof.} Recently related theoretical
work on enhanced diffusion was published \cite{latora}

{\bf Acknowledgment} We thank A. Aharony, P. Levitz, I. Sokolov and R. Metzler
for helpful discussions.

\subsection{Appendix A}

As mentioned we use a lattice model for
the simulation so that $\xi$ is an integer.
We use the transformation
\begin{equation}
\xi=\mbox{INT}\left\{ \left[ \tan\left( { u \pi \over 2} \right) \right]
^{1 / \gamma} \right\}+1.
\label{eqAP1}
\end{equation}
Here $\mbox{I}=\mbox{INT}\{ z  \}$ is the integer
closest to $z$ satisfying $I \le z$.
In Eq. (\ref{eqAP1})
$u$ is a random variable distributed uniformly
according to
\begin{equation}
 0\le u_{min} \le u \le u_{max} \le 1,
\label{eqAP1a}
\end{equation}
where $u_{min}$ and $u_{max}$ are cutoffs. It is easy
to generate the random variable $u$ on a computer.
The probability to find an interval of length $\xi$ is,
\begin{equation}
\mu\left( \xi \right)= \int_{\xi-1}^{\xi} \mu_c(y) dy
\label{eqAP2}
\end{equation}
with
\begin{equation}
\mu_c(y) = \left\{
\begin{array}{ccc}
0 &  \ \  y< y_{min} \\
 \ &  \ \   \\
 { 2 \gamma \over \pi \Delta } { y^{\gamma - 1} \over 1 + y^{2 \gamma}} &  \ \
y_{min}< y < y_{max} \\
 \ &  \ \   \\
 0  &  \ \ y > y_{max}. \\
\end{array}
\right.
\label{eqAP3}
\end{equation}
Here
\begin{equation}
y_{min} = \left[ \tan\left( {u_{min} \pi \over 2} \right)\right]^{1/ \gamma},
\ \ \ \
y_{max} = \left[ \tan\left( {u_{max} \pi \over 2} \right)\right]^{1/ \gamma}
\label{eqAP4}
\end{equation}
are the cutoffs of $\mu_c(y)$. When $u_{min}=0$  and $u_{max}=1$ we
have $y_{min}=0$ and $y_{max}= \infty$.
In Eq. (\ref{eqAP3}) $\Delta=u_{max} - u_{min}$ determines the normalization
condition $\int_{0}^{\infty} \mu_c(y) dy = 1$.

 To derive Eqs. (\ref{eqAP2})-(\ref{eqAP4}) we use the transformation
$ y = \left[ \tan\left( u \pi / 2 \right) \right]^{1/ \gamma}$,
and then $\mu_c\left( y \right) = g\left( u \right) | du/dy|$,
where $g\left( u \right)$ is the uniform probability density
 of $u$.

 For large $\xi$ we find
\begin{equation}
\mu\left(\xi \right) \sim \left\{
\begin{array}{ccc}
{ 2 \gamma \over \pi \Delta}\xi ^{ - 1 - \gamma} & \ \ \xi< \xi_{max} \\
 \ &  \ \   \\
0                                               & \ \ \xi > \xi_{max},
\end{array}
\right.
\label{eqP5}
\end{equation}
with
$\xi_{max} = \mbox{INT} \left\{ \left[ \tan\left( u_{max} \pi/2 \right) \right]^
{1 / \gamma} \right\} + 1$.
When $u_{max}=1$ the
second moment of $\mu(\xi)$ diverges.

 In our numerical simulations we consider
$u_{min}=1/2$, $u_{max}=1$ and $\gamma=3/2$. Then
$ \langle  \xi \rangle \simeq 4.031$ and for large $\xi$,
we find 
\begin{equation}
\mu(\xi)\sim (6/ \pi) \xi^{- 5/2} \ \ \ \ \ \ \ \xi>>1.
\label{eqEX01}
\end{equation}

\subsection{Appendix B}

 The calculation of $\hat{Q}_b(u)$ can be found in
\cite{Feller,COX}. The probability that 
the interval $(0,t)$ is empty is
\begin{equation}
G_0(t)=1-\int_0^t h(\tau) d \tau 
\label{eqAPB1}
\end{equation}
and in Laplace space $\hat{G}_0(u)=[1 - \hat{h}(u) ] / u$.
The Laplace transform of $G_r(t)$ for $r\ge 1$
is found using convolution
\begin{equation}
\hat{G}_r(u) = \hat{h}(u) \hat{\mu}^{r - 1}(u) \hat{W}(u).
\label{eqAPB2}
\end{equation}
$W(t)=1- \int_0^{t} \mu(\xi) d \xi$ is the probability 
that an interval of length $(0,t)$ is empty, given that
a scatterer occupies $0^-$. In Laplace space $\hat{W}(u)=[1- \hat{\mu}(u)]/u$.
Using Eqs. (\ref{eqAPB1}), (\ref{eqAPB2})
and (\ref{eqLL07})
we find (\ref{eqLL08}).

\end{document}